\documentclass[aps,prb,twocolumn,showpacs,draft,floatfix]{revtex4}
\usepackage{amsbsy}
\usepackage{amsmath}
\usepackage{epsf}
\usepackage{dcolumn}
\def\c{\hbox{const}}
\def\d{\mathop{\hbox{d}}\!}
\def\nab{\hbox{$\boldsymbol{\nabla}$}}
\def\hphi{\hbox{$\boldsymbol{\hat\phi}$}}
\def\mp{m_{\parallel}}
\def\mo{m_{\perp}}

\def\Do{{\bf D}_{\perp}}
\def\alphaT{\alpha_T}
\def\alphaTm{\alpha_{T,m}}

\begin{document}

\title{Correlation lengths for vortex-liquid freezing\\
       in a model of layered high-temperature superconductor}
\author{A.~Zamora}
\email{zamora@nuclecu.unam.mx}
\affiliation{Instituto de Ciencias Nucleares, UNAM, Apdo. Postal 70-543, M\'exico D.F. 04510}
\date{\today}

\begin{abstract}
We present results from extensive numerical simulations on the 
Lawrence-Doniach model within the lowest Landau level approximation 
in the unconventional spherical geometry. We study the in-layer 
pancake vortex density-density correlation function (intra-layer 
structure factor) and the layer-to-layer order-parameter correlation 
function along the direction perpendicular to the layers. Our results 
show strong evidence for the existence of a single first-order phase 
transition at which both inter-layer coupling and appearance of 
crystalline order in the in-layer vortex correlations take place as 
temperature is lowered.
\end{abstract}


\maketitle

The study of the phase diagram of layered superconductors in the 
presence of a magnetic field has attracted much attention since 
the discovery of the high-temperature superconducting cuprates%
~\cite{p:Bednorz&-1986}. In particular, the most interesting 
effects are found in the mixed phase~\cite{p:Shubnikov&-1937} 
where the external magnetic field penetrates the sample in the 
form of quantized flux-lines (or vortices). Within the mixed 
phase, originally described by Abrikosov~\cite{p:Abrikosov-1957} 
as a vortex lattice, a high-temperature phase with finite resistivity 
has been found and called a vortex liquid. While the vortex-liquid 
phase is quite narrow in conventional type-II superconductors, its 
width is known to become large in highly anisotropic cuprates%
~\cite{p:Blatter&-1994} (e.g. Bi$_2$Sr$_2$CaCu$_2$O$_8$.) Assuming a 
specimen completely free of defects, the low-temperature phase is 
expected to be the Abrikosov vortex-lattice and the phase boundary 
separating the vortex liquid and vortex lattice appears to be a line 
of first-order phase transitions usually called the melting line%
~\cite{p:Hwang&-1996} (which essentially coincides with the magnetic 
irreversibility-line~\cite{p:Pastoriza&-1994}). Results from 
experiments, analytic calculations, and numerical simulations on 
several models of layered superconductors have shown that, in 
addition or simultaneous to melting, there is a transition 
associated with layer decoupling%
~\cite{p:Shibauchi&-1999,p:Colson&-2003,p:Qiu&-2000,
p:Glazman&-1991,p:Rodriguez&-2002,p:Hu&-1997,p:Mohler&-2000}. 
In another simulation, Wilkin and Jensen~\cite{p:Wilkin&-1997} on 
a layered model find a first-order transition associated with 
layer decoupling followed by melting but which shows no obvious
thermodynamic signature. An alternative scenario was proposed by
Kienappel and Moore~\cite{p:Kienappel&-1999}. From 
Langevin-dynamics simulations on the Lawrence-Doniach (LD) model 
within the lowest Landau level (LLL) approximation and in the 
same spherical (S) geometry that we consider here (we shall call 
this the S-LD-LLL model) they find hysteretic effects on several 
quantities which disappear on increasing the interlayer 
coupling-strength. They attribute this result to a line of first-order 
transitions ending at a critical point which separates phases of 
coupled and decoupled vortex-liquid (expected to freeze only at 
zero temperature.)

In this paper we present an alternative view based on Monte Carlo
(MC) simulations in the same S-LD-LLL model. From studies of inlayer 
vortex-density correlations as well as layer-to-layer order parameter 
correlations we find strong evidence for vortex-liquid freezing 
(simultaneous to inter-layer coupling) at a finite temperature. As in 
Ref.~\cite{p:Kienappel&-1999}, we model this layered superconductor by 
a system of $M$ concentric spherical-layers of thickness $d_0\ll R$, 
where $R$ is the average radius of the sample, and interlayer spacing 
$s\ge d_0$. The reason for the choice of this geometry, instead of 
the usual array of plane-layers, 
is that this geometry guarantees full rotational and translational
invariance of the vortex system (which quasi periodic boundary 
conditions in the planar geometry do not). This allows for the 
in-layer free movement of vortices without any boundary constraint, 
which is expected to better describe a real superconducting sample.
This has been previously discussed in detail for the
single-layer case by Dodgson and Moore~\cite{p:Dodgson&-1997}. An
external magnetic field ${\bf H}({\bf r})=(H_0 R^2/r^2){\hat{\bf r}}$
[with $H_0$ a constant field], orthogonal 
to all layers in every point, is generated by an infinitely long and 
thin solenoid ending at the center of the concentric spheres. We neglect 
the fluctuations in the magnetic induction inside the sample so that it 
has the mean value $B=\mu_0 H(R)$ across all superconducting layers. 
Flux quantization requires that the magnetic induction penetrating 
the sample be such that $4\pi R^2 B=N\Phi_0$, where $\Phi_0=h/2|e|$ 
is the flux quantum and $N$ the number of flux lines or vortices. 
Our choice of gauge that satisfies ${\bf B}=\nab\times{\bf A}$ is 
${\bf A}=(\mu_0 H_0 R^2/r)\tan(\theta/2){\hphi}$. We assume that the 
system without impurities accepts a description in terms of the 
LD Hamiltonian functional~\cite{p:Lawrence&-1971} which, under the 
previous assumptions, can be written as
\begin{eqnarray}
\label{eq:clean-H-functional}
&&{\cal H}[\psi_n({\bf r})]=\sum_{n=1}^M
d_0\int\d^2 r\left[\alpha(T)|\psi_n|^2+\frac{\beta}{2}|\psi_n|^4+{}
\right.\qquad\nonumber\\
&&\quad\qquad\left.{}+\frac{1}{2\mo}|\Do\psi_n|^2
+\frac{\hbar^2}{2\mp s^2}|\psi_{n+1}-\psi_n|^2\right].
\end{eqnarray}
Here $\psi_n({\bf r})$ is the two-dimensional Ginzburg-Landau (GL) 
order parameter in layer $n$ and $\Do=-i\hbar\nab_{\perp}-2e{\bf A}$ 
the gauge covariant derivative operator (acting on the surface of 
the sphere). $\alpha(T)$ and $\beta$ are the usual parameters from 
the GL theory~\cite{p:Landau&-1950}. $\mo$ and $\mp$ are the effective 
Cooper-pair masses in the perpendicular and parallel directions relative 
to the radial magnetic field. The LLL approximation consists in expanding 
the GL order parameter in each layer as a linear combination of the 
degenerate eigenfunctions of $D_{\perp}^2=\Do\cdot\Do$ corresponding to 
the lowest eigenvalue $2|e|\hbar\mu_0 H$. An orthonormal set of functions 
for the LLL on the surface of the sphere is~\cite{p:Roy&-1983}
\begin{equation}
\label{eq:LLL}
\varphi_{q,N}(\theta,\phi)={\cal N}_{q,N} e^{iq\phi}\sin^q(\theta/2)
\cos^{N-q}(\theta/2),
\end{equation}
where $q=0,1,\dots,N$ labels the degeneracy of the LLL. The normalization
factor is ${\cal N}_{r,s}=[(s+1)C_s^r/4\pi R^2]^{1/2}$, with 
$C_s^r=s!/r!(s-r)!$ the binomial coefficient. We expand the superconducting 
order parameter in layer $n$ as 
$\psi_n({\bf r})=Q\sum_{q=0}^N v_{n,q}\varphi_{q,N}(\theta,\phi)$ with 
$Q=(\Phi_0 k_B T/\beta d_0 B)^{1/4}$ and measure lengths in units of 
the magnetic length $l_m=(\Phi_0/2\pi B)^{1/2}$. In these units 
$R=(N/2)^{1/2}l_m$.

Within the LLL approximation, the Hamiltonian functional in Eq. 
(\ref{eq:clean-H-functional}) becomes a function of the complex 
coefficients $\{v_{n,q}\}\equiv{\bf v}$ given by
\begin{eqnarray}
\label{eq:LLL-clean-H-func}
&&{\cal H}({\bf v})/k_B T=\sum_{n=1}^M\left[
\alphaT\sum_{q=0}^N |v_{n,q}|^2 +
\frac{1}{2N}\sum_{p=0}^{2N} |U_{n,p}|^2 +{}
\right.\nonumber\\
&&\qquad\qquad\qquad\qquad\left.{}+\eta|\alphaT|\sum_{q=0}^N 
|v_{n+1,q}-v_{n,q}|^2\right],
\end{eqnarray}
where the effective temperature-field parameter is~\cite{c:alphat}
\begin{equation}
\label{eq:alphaT}
\alphaT=\frac{d_0 Q^2}{k_B T}\left(\alpha(T)+\frac{|e|\hbar\mu_0 H}{\mo}
\right).
\end{equation}
[Note that $\alphaT=0$ corresponds to the mean-field $H_{c2}(T)$ line
and $\alphaT=-\infty$ corresponds to $T=0$.] The parameter that 
determines the coupling strength between adjacent layers is then just 
$\eta=\hbar^2/(2\gamma^2 s^2|eB\hbar+\mo\alpha(T)|)$, where
$\gamma=(\mp/\mo)^{1/2}$ is the anisotropy parameter. 
In the second term of Eq. (\ref{eq:LLL-clean-H-func}) we have used
\begin{eqnarray}
\label{eq:U_pn}
&&U_{n,p}=\left(\pi^{1/2}N/R{\cal N}_{p,2N}\right)\sum_{q=0}^N 
{\cal N}_{q,N} {\cal N}_{p-q,N} v_{n,q} v_{n,p-q}
\quad\nonumber\\
&&\qquad\qquad\qquad\qquad\qquad\quad\times\Theta(p-q)\Theta(N-p+q),
\end{eqnarray}
with $\Theta(x)$ the Heaviside step function, equal to zero for $x<0$ 
and one for $x\ge 0$. The Hamiltonian describing our model of a layered 
superconductor within the LLL approximation is that in Eq. 
(\ref{eq:LLL-clean-H-func}). Note that ${\cal H}$ is a function of the 
complex variables $\{v_{n,q}\}$ and depends on the temperature, 
$\alphaT$, and layer-coupling, $\eta$, parameters.

The equilibrium properties of this system are determined by its partition 
function
\begin{equation}
\label{eq:Partition-Function}
Z(T,H)=\int\prod_n{\cal D}\psi_n{\cal D}\psi_n^{*}{\cal D}{\bf A}
\exp(-{\cal H}[\psi_n,{\bf A}]/k_B T),
\end{equation}
whose value is basically controlled by the order parameter configurations
that minimize the Hamiltonian functional ${\cal H}$. In the LLL approximation,
which neglects fluctuations in the magnetic induction and restricts the
order parameter (in each layer) to the set of functions spanning the LLL,
the partition function can be written as
\begin{equation}
\label{eq:LLL-Partition-Function}
Z(T,H)=\int\prod_{n,q}\d v_{n,q}\d v^{*}_{n,q}\exp[-{\cal H}({\bf v})/k_B T].
\end{equation}
Thermal averages of the quantities of interest $X(T,H)$,
\begin{eqnarray}
\label{eq:LLL-Thermal-Av}
&&\langle X(T,H) \rangle = \frac{1}{Z(T,H)}
\int\prod_{n,q}\d v_{n,q}\d v^{*}_{n,q}X({\bf v})
\qquad\nonumber\\
&&\qquad\qquad\qquad\qquad\qquad\qquad\times\exp[-{\cal H}({\bf v})/k_B T],
\end{eqnarray}
are calculated by means of MC simulations using the standard
Metropolis algorithm~\cite{p:Metropolis&-1953}. We calculate thermal 
averages by the Importance Sampling Method~\cite{p:Binder-1997}.

The physical quantities that we have focused on to examine in-layer 
vortex correlations are the structure factor (in each layer) defined 
as~\cite{b:Chaikin&-1997}
\begin{equation}
\label{eq:structure-factor}
S({\bf k}) = \frac{1}{N}\left\langle\sum_{i,j=1}^N
{\rm e}^{-i{\bf k}\cdot({\bf x}_i-{\bf x}_j)}\right\rangle,
\end{equation}
where $\{{\bf x}_l\}$ are the positions of the $N$ pancake vortices 
(on that layer) and we have chosen to parametrize the wave vector 
in polar coordinates as ${\bf k}=(k_x,k_y)=(k\cos\phi,k\sin\phi)$,
and the rotationally averaged structure function (also in each 
layer) given by
\begin{equation}
\label{eq:structure-function}
\Delta(k) = \frac{1}{2\pi}\int_0^{2\pi}{\rm d}\phi S({\bf k}). 
\end{equation}
A Lorentzian fit to the first peak of this function is in agreement 
with an exponential decay of vortex density-correlations in real 
space~\cite{p:Dodgson&-1997}. The length scale governing that decay, 
$\xi_D$, is proportional to the inverse-width at half peak, 
$\delta^{-1}$, of the Lorentzian.

\begin{figure}
\centerline{\epsfxsize= 2.7cm\epsfbox{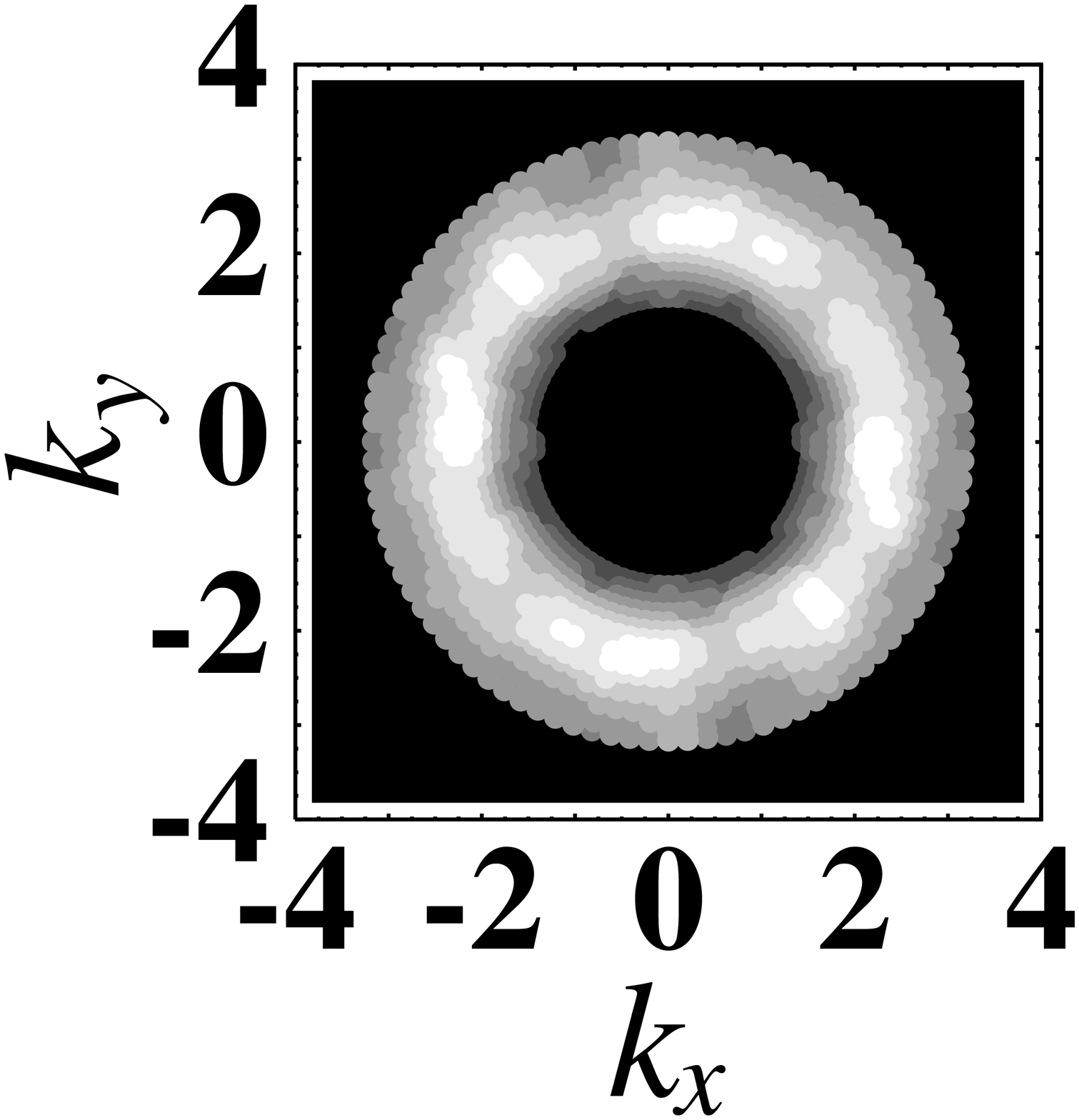}\,\,%
            \epsfxsize= 2.7cm\epsfbox{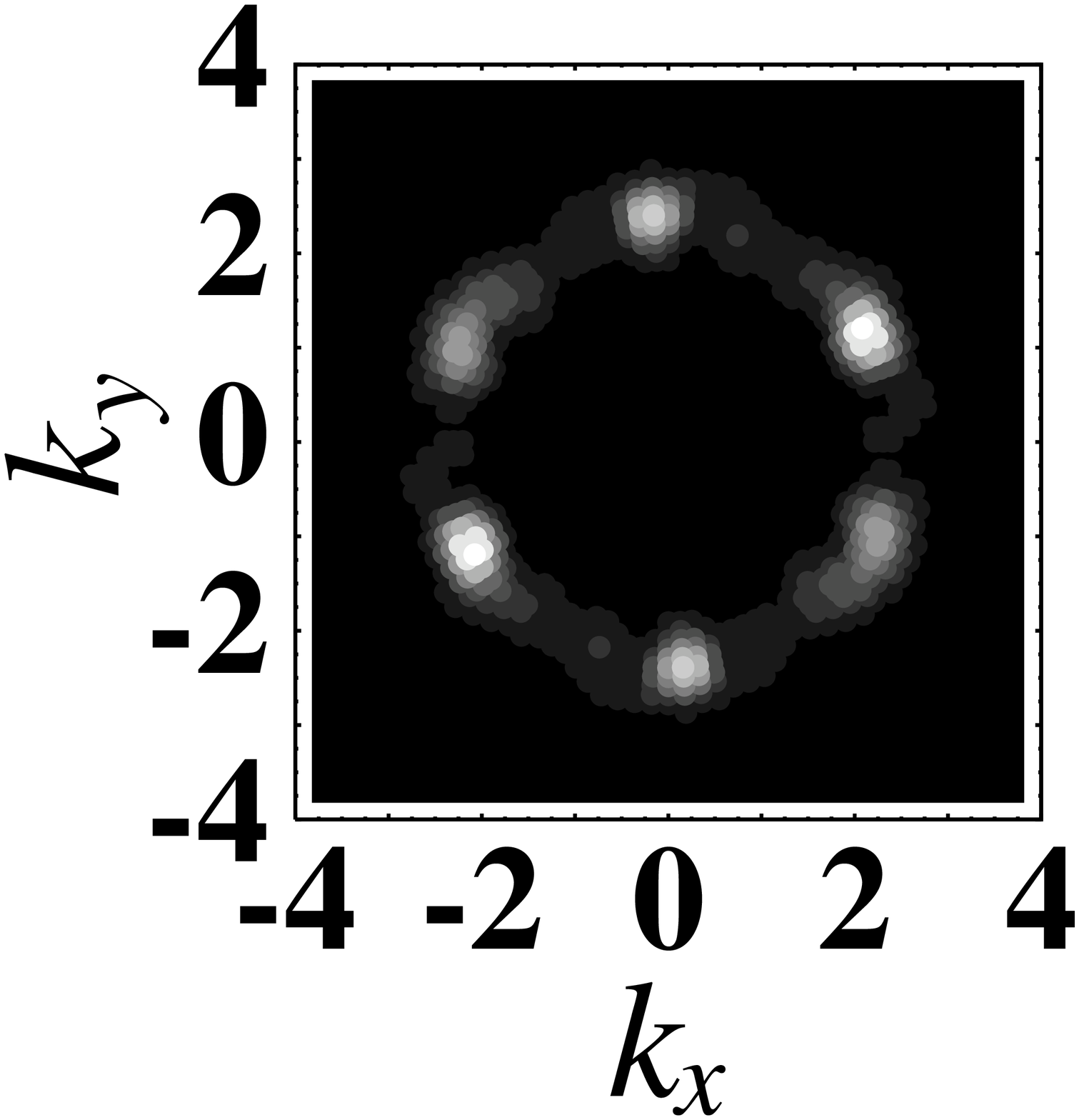}\,\,%
            \epsfxsize= 2.7cm\epsfbox{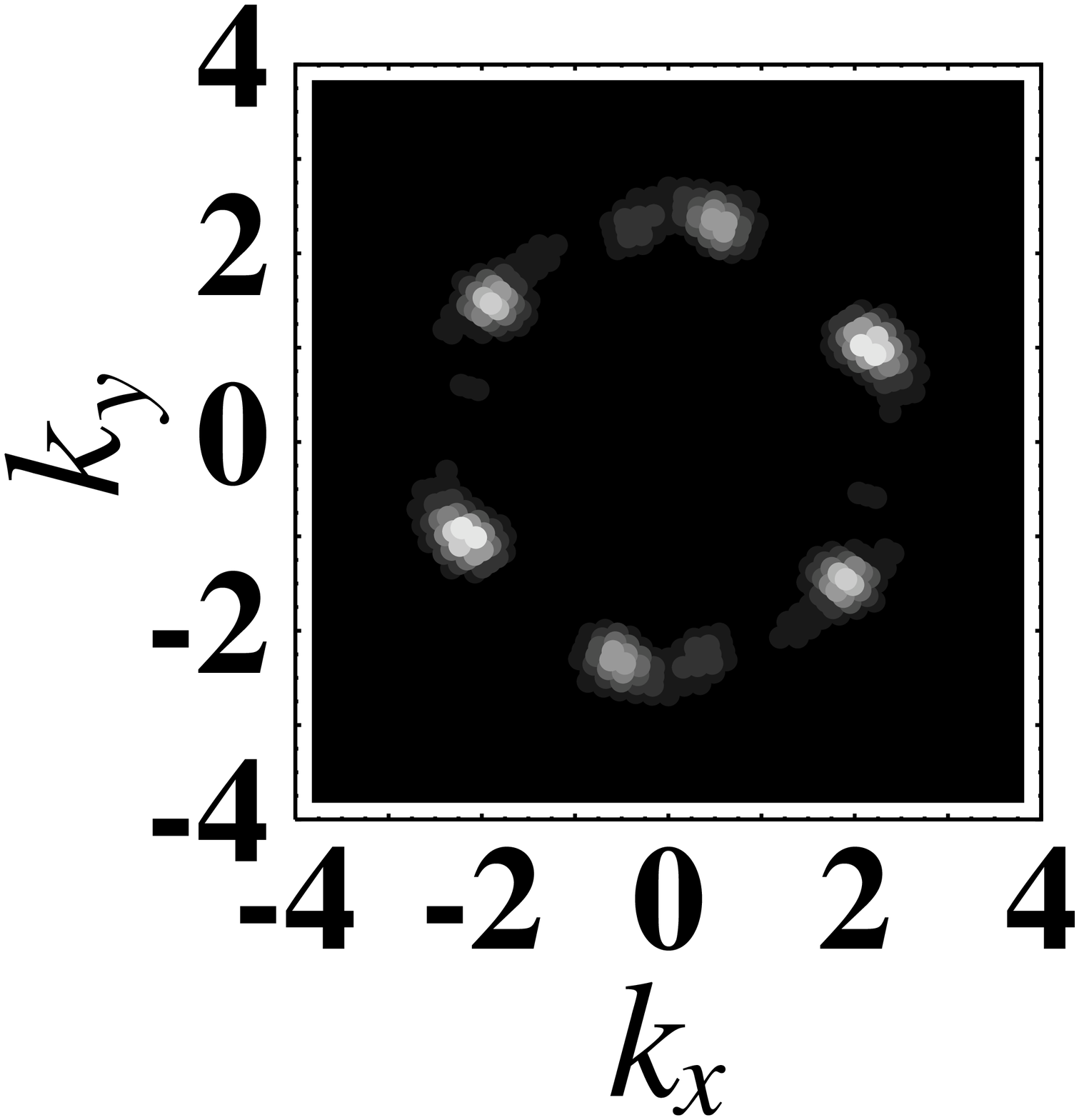}}
  \vglue 0.0em
  \centerline{\hglue 2.05cm (a) \hglue 2.28cm (b) 
              \hglue 2.28cm (c) \hglue 1.34cm}
  \vglue 0.1em
\centerline{\epsfxsize= 2.7cm\epsfbox{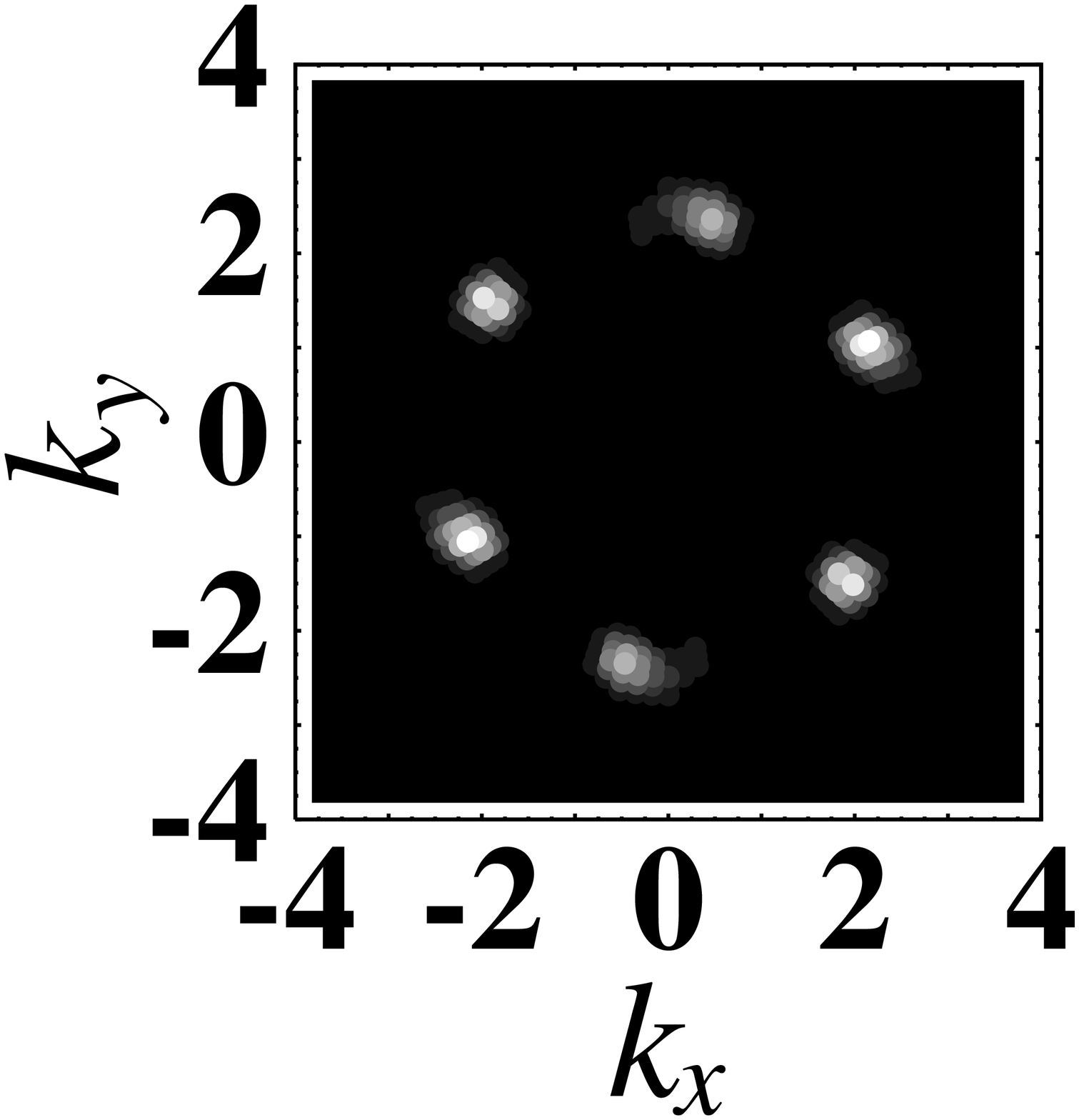}\,\,%
            \epsfxsize= 2.7cm\epsfbox{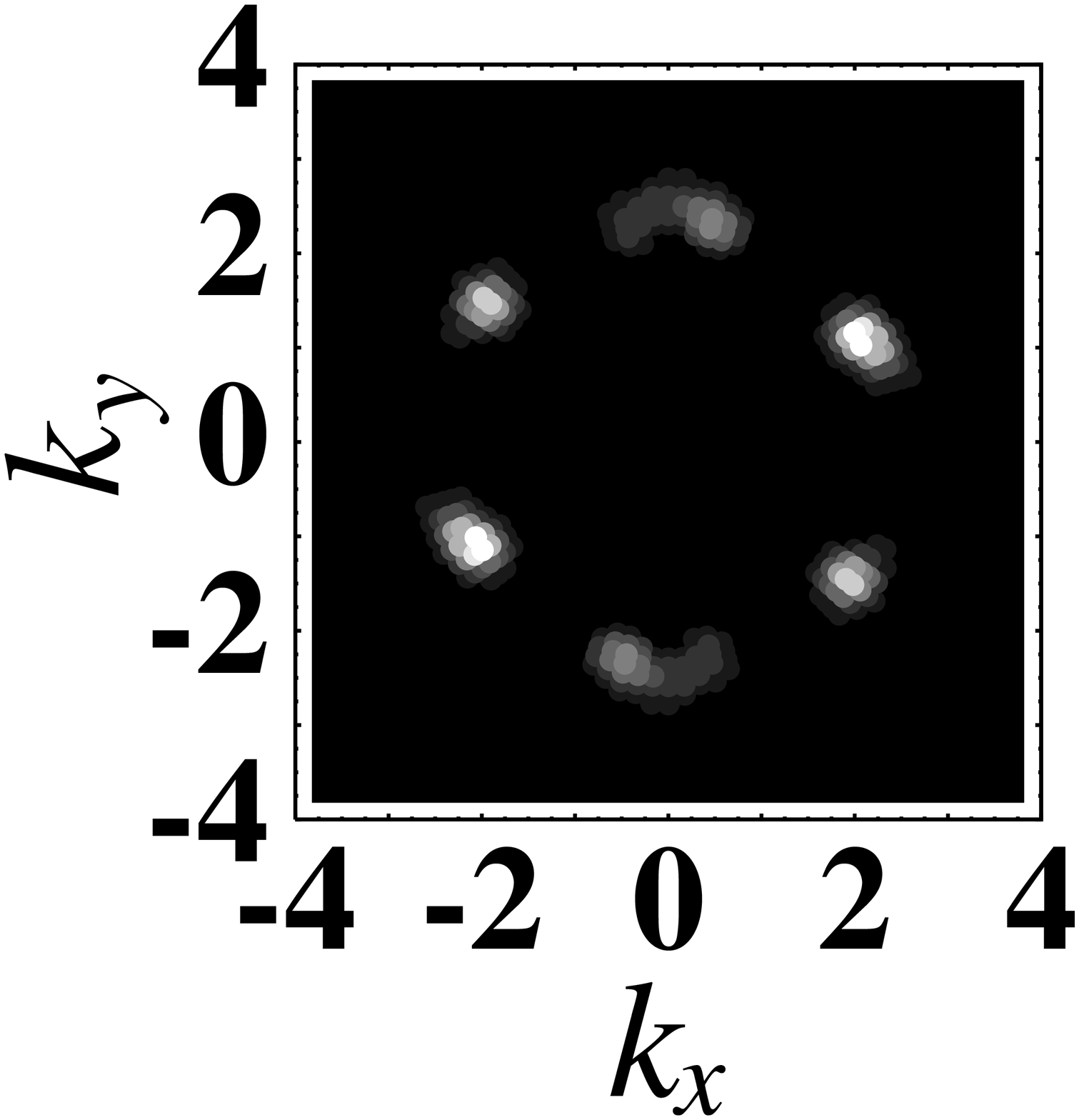}\,\,%
            \epsfxsize= 2.7cm\epsfbox{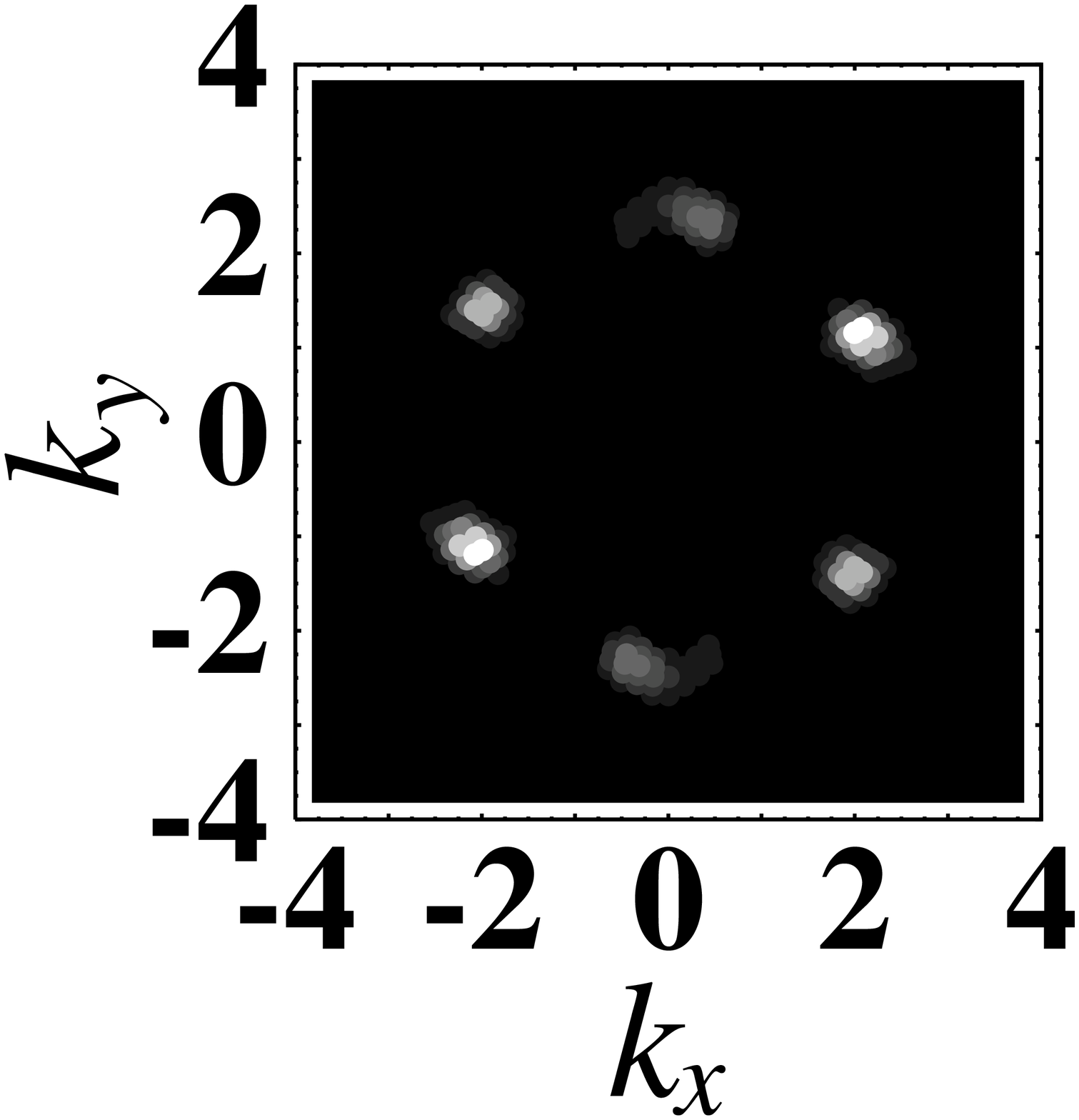}}
  \vglue 0.0em
  \centerline{\hglue 2.00cm (d) \hglue 2.29cm (e) 
              \hglue 2.29cm (f) \hglue 1.32cm}
  \vglue 0.1em
  \caption{Structure factor, $S({\bf k})$, at layer $M/2$ of the clean 
  layered superconductor with interlayer coupling $\eta=0.14$ for $(N,M)=(72,18)$ 
  as the temperature is lowered: $\alphaT=-3,-4,-5,-6,-7,-8$. 
  [(a),(b),(c),(d),(e),(f) respectively.] The central maximum at ${\bf k}={\bf 0}$
  has been removed for clarity. Note the appearance of Bragg peaks
  at a temperature between $\alphaT=-3$ and $-4$.}
  \label{Fig:S(k)}
\end{figure}

\begin{figure}
\centerline{\epsfxsize= 8.3cm\epsfbox{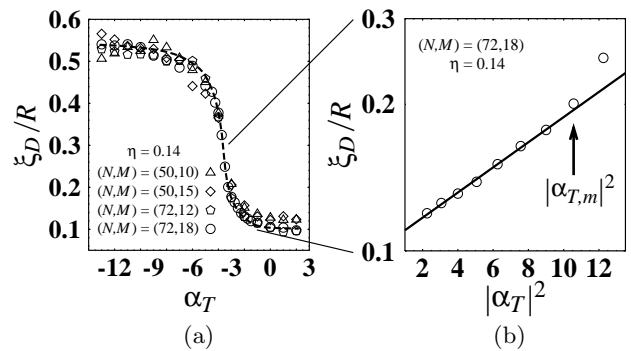}}
  \vglue 0.0em
  \centerline{\hglue 5.20cm (a) \hglue 3.60cm (b) \hglue 4.18cm}
  \vglue 0.1em
  \caption{In-layer density correlations at layer $M/2$ of the clean 
  layered-HTSC as the temperature is lowered. (a) Scaled vortex-density 
  correlation-length, $\xi_D/R$, for different system sizes at $\eta=0.14$. 
  The dashed curve is just a guide to the eye. (b) The same characteristic 
  length-scale, $\xi_D/R$, as a function of $|\alphaT|^2$ in log-linear 
  scale for the system $(N,M)=(72,18)$ with exponential fit (solid line). 
  [The arrow marks the transition temperature for $\eta=0.14$, 
  $\alphaTm=-3.25$.]}
  \label{Fig:Xi(alphat)}
\end{figure}

Numerical measurements of these quantities have been made for runs up 
to $1.92\times 10^6$ MC steps~\cite{c:MC-steps} with systems as big 
as 18 layers containing 72 vortices per layer. We have studied the range 
of effective temperatures $-13\le\alphaT\le 2$ (while cooling) for inter-layer 
couplings $\eta=0.14,0.5,1,2,10,100$. One of the most notable features 
of a freezing transition, which is the appearance of Bragg peaks in the 
structure factor $S({\bf k})$, is observed at an effective temperature
$\alphaT\simeq -4$ (for $\eta=0.14$) and at higher temperatures for
$\eta\ge 0.5$. Fig. \ref{Fig:S(k)} shows this structure factor as 
temperature is lowered (see caption). Fig. \ref{Fig:S(k)}(b) corresponds 
to the temperature $\alphaT=-4$ at which the vortex system is essentially 
crystallized. We notice that at and below $\alphaT=-5$ (Figs. 
\ref{Fig:S(k)}(c) to \ref{Fig:S(k)}(f)) the Bragg peaks appear exactly
at the same positions. 
This is consistent with a legitimate crystalline vortex-phase for these 
low temperatures.
Another signature of the vortex-liquid freezing is seen in the rapid 
growth of vortex density-correlations, $\xi_D$, below $\alphaT=-3$ (for 
the same coupling $\eta=0.14$). This is shown in Fig. \ref{Fig:Xi(alphat)}, 
where the scaled density correlation length $\xi_D/R$ is plotted. Fig. 
\ref{Fig:Xi(alphat)}(a) shows the abrupt increase of this length scale 
as a function of our temperature parameter $\alphaT$ for all system sizes 
studied. Fig. \ref{Fig:Xi(alphat)}(b) shows the same characteristic 
length-scale, $\xi_D/R$, plotted (in log-linear scale) against $|\alphaT|^2$. 
We observe that an exponential fit (straight-line there) to the data is 
appropriate at high temperatures (i.e. at low values of $|\alphaT|^2$), but 
breaks down in the low temperature regime where correlations grow even faster 
than that. The exponential behavior, $\xi_D\sim\exp(\c|\alphaT|^2)$, has 
been predicted by Moore~\cite{p:Moore-1997} in connection with bulk 
anisotropic-superconductors in the continuum limit at their lower critical 
dimension. In that investigation, however, this rapid growth of correlations 
is not attributed to a thermodynamic phase transition but just to a crossover. 
Returning to Fig. \ref{Fig:Xi(alphat)}(b), we remark that the very last points 
mark the onset of in-layer vortex correlations growing even faster than 
$\xi_D\sim\exp(\c|\alphaT|^2)$ [compare with Fig. \ref{Fig:Xi(alphat)}(a)
for lower temperatures.] We believe this can occur because a thermodynamic 
first-order transition takes place at the temperature $\alphaTm$ where 
$\xi_D$ departs from the straight line in Fig. \ref{Fig:Xi(alphat)}(b). 
This takes place at $\alphaTm=-3.25$ for $\eta=0.14$. In fact, comparison
with the work of Hu and MacDonald~\cite{p:Hu&-1997} gives excellent
agreement for the melting (freezing) transition at these values.

\begin{figure}
\centerline{\epsfxsize= 2.7cm\epsfbox{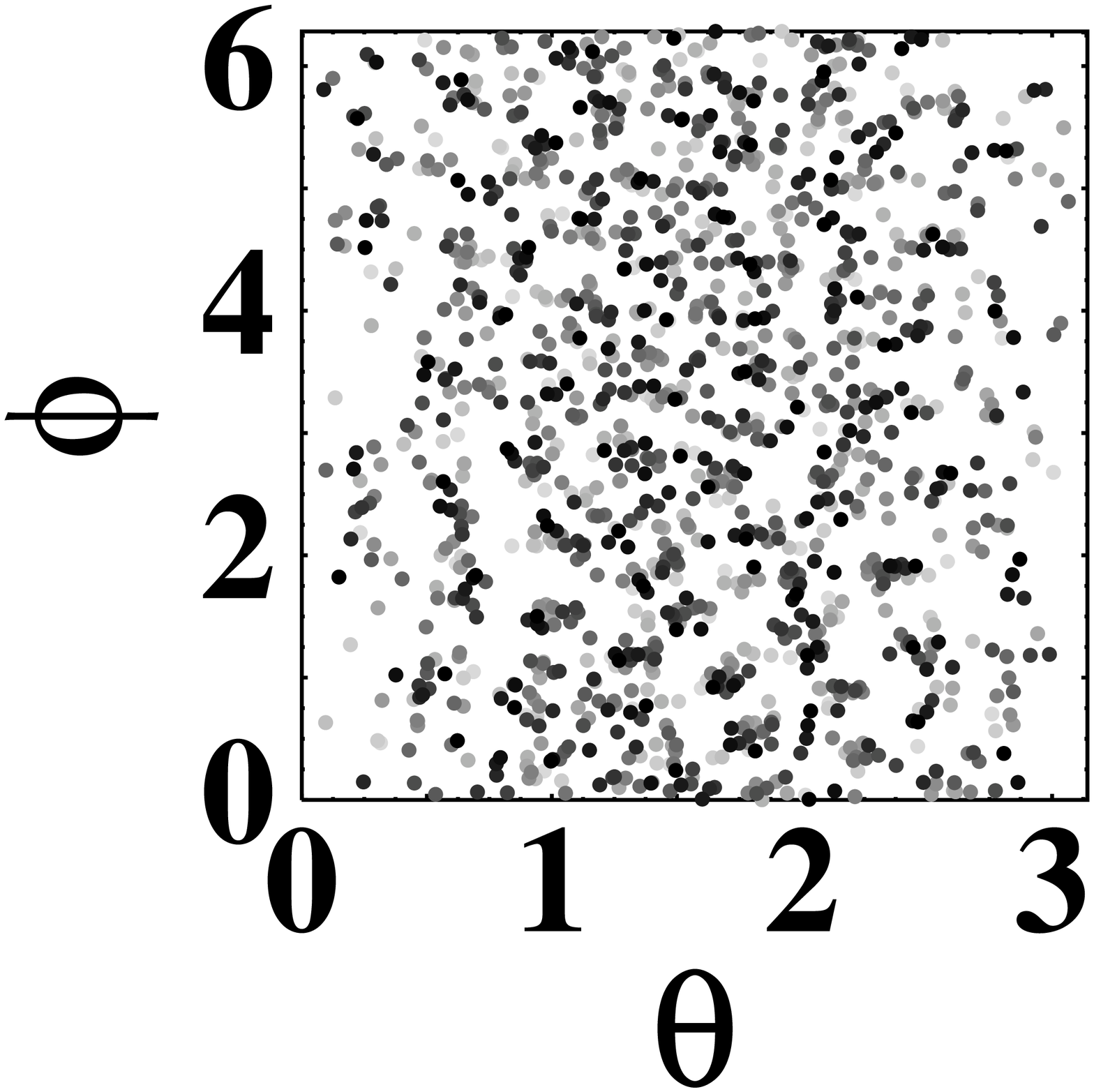}\,\,\,%
            \epsfxsize= 2.7cm\epsfbox{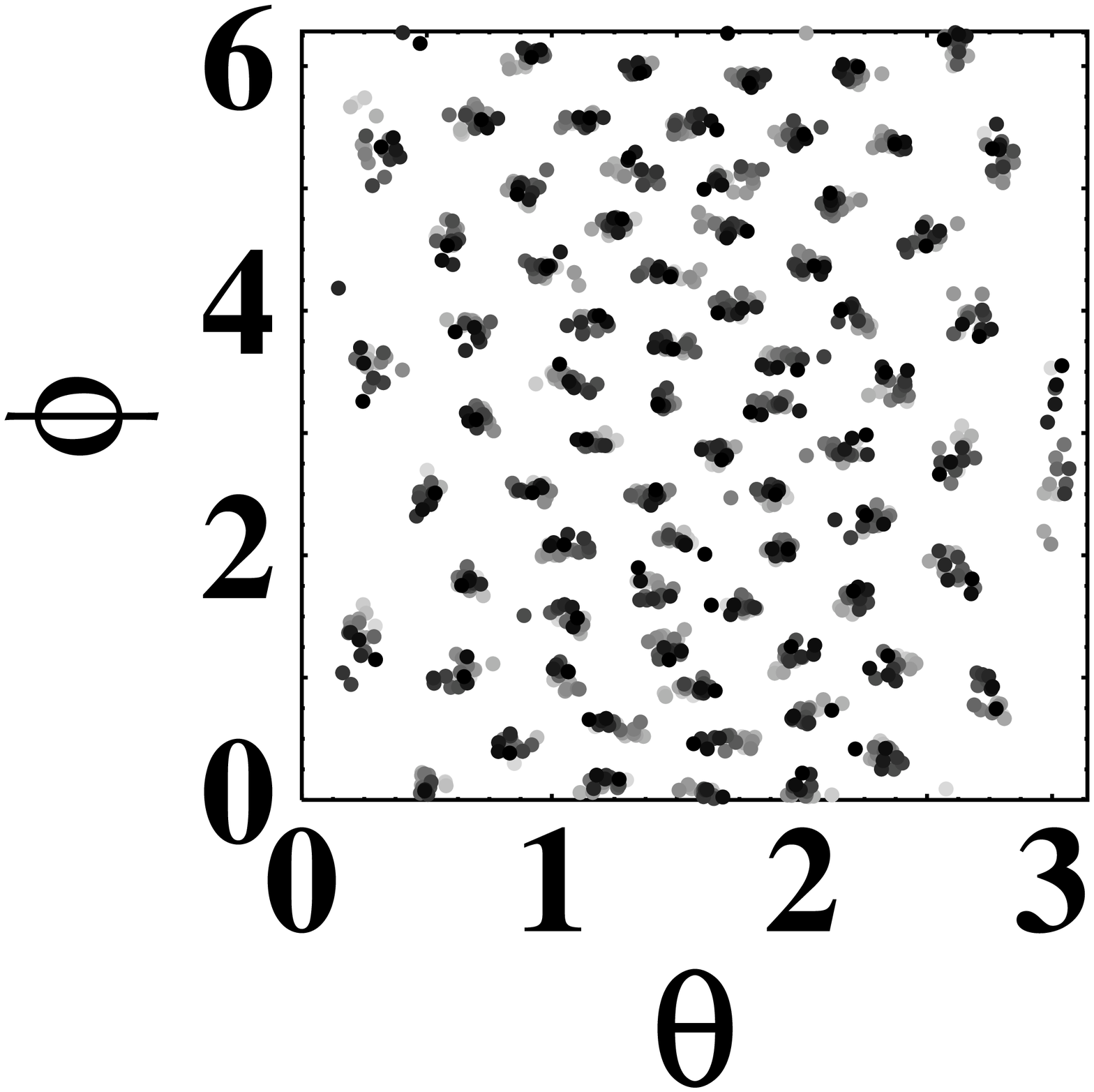}\,\,\,%
            \epsfxsize= 2.7cm\epsfbox{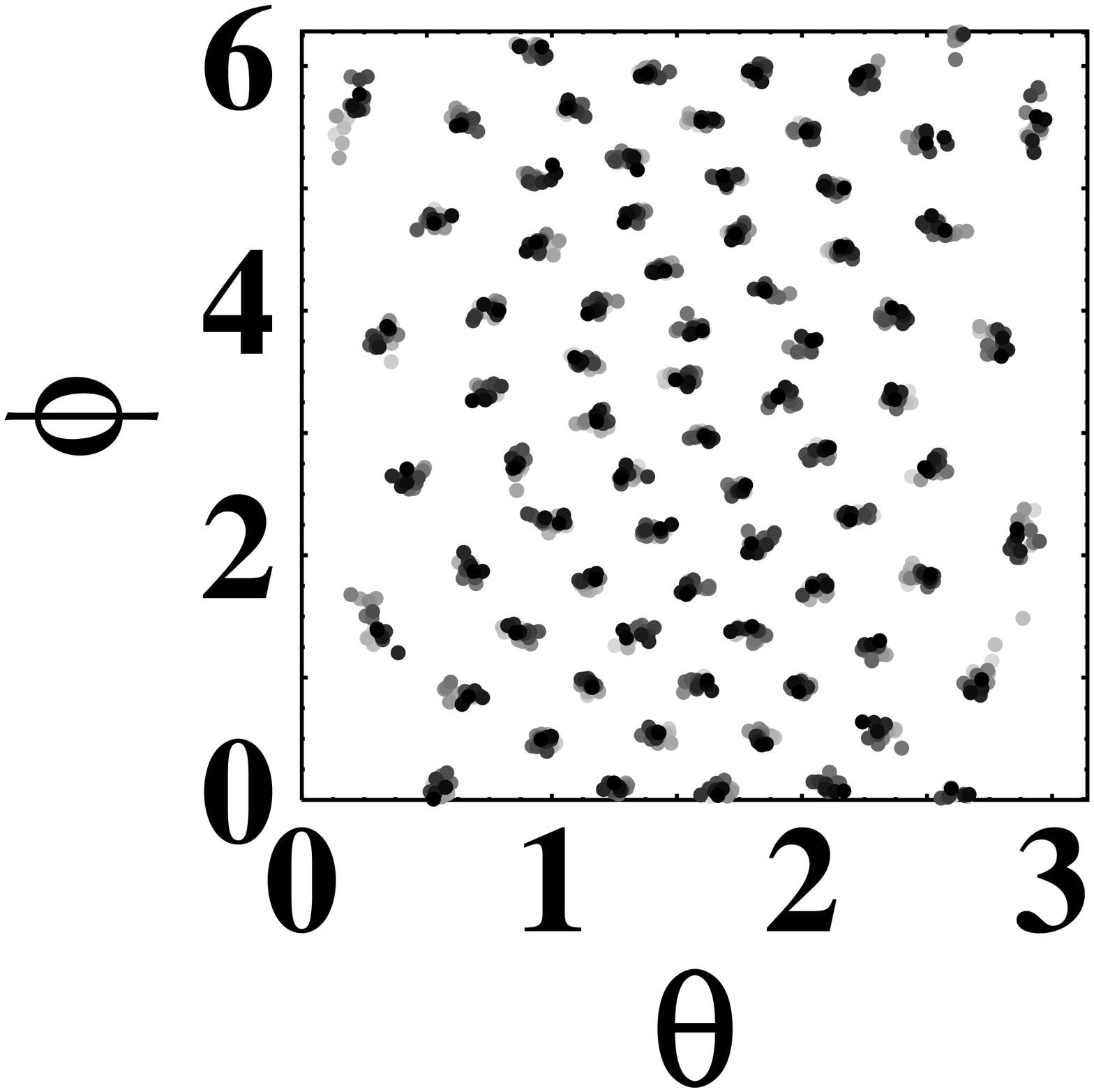}}
  \vglue 0.0em
  \centerline{\hglue 1.97cm (a) \hglue 2.35cm (b) 
              \hglue 2.35cm (c) \hglue 1.20cm}
  \vglue 0.1em
\centerline{\epsfxsize= 2.7cm\epsfbox{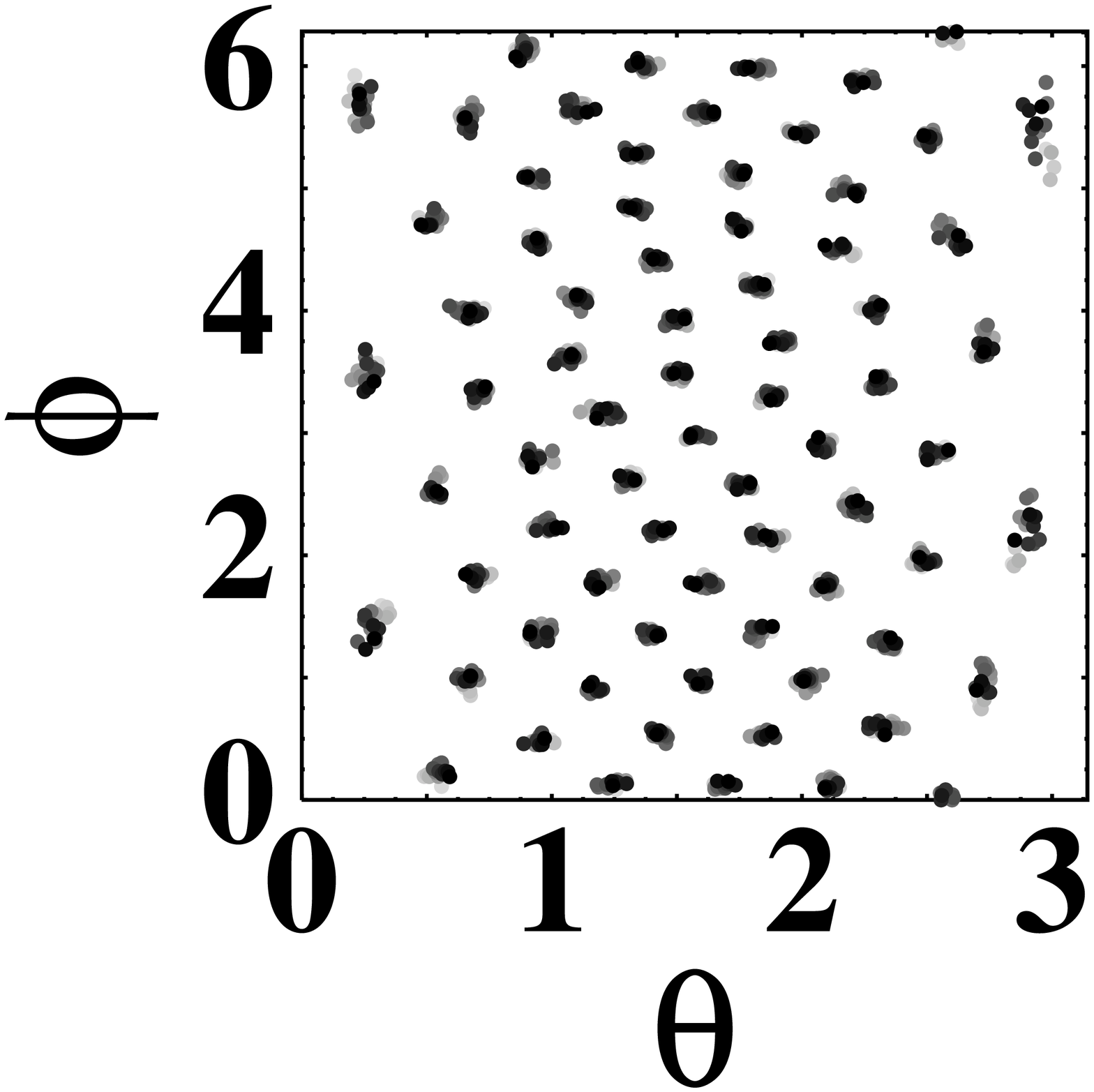}\,\,\,%
            \epsfxsize= 2.7cm\epsfbox{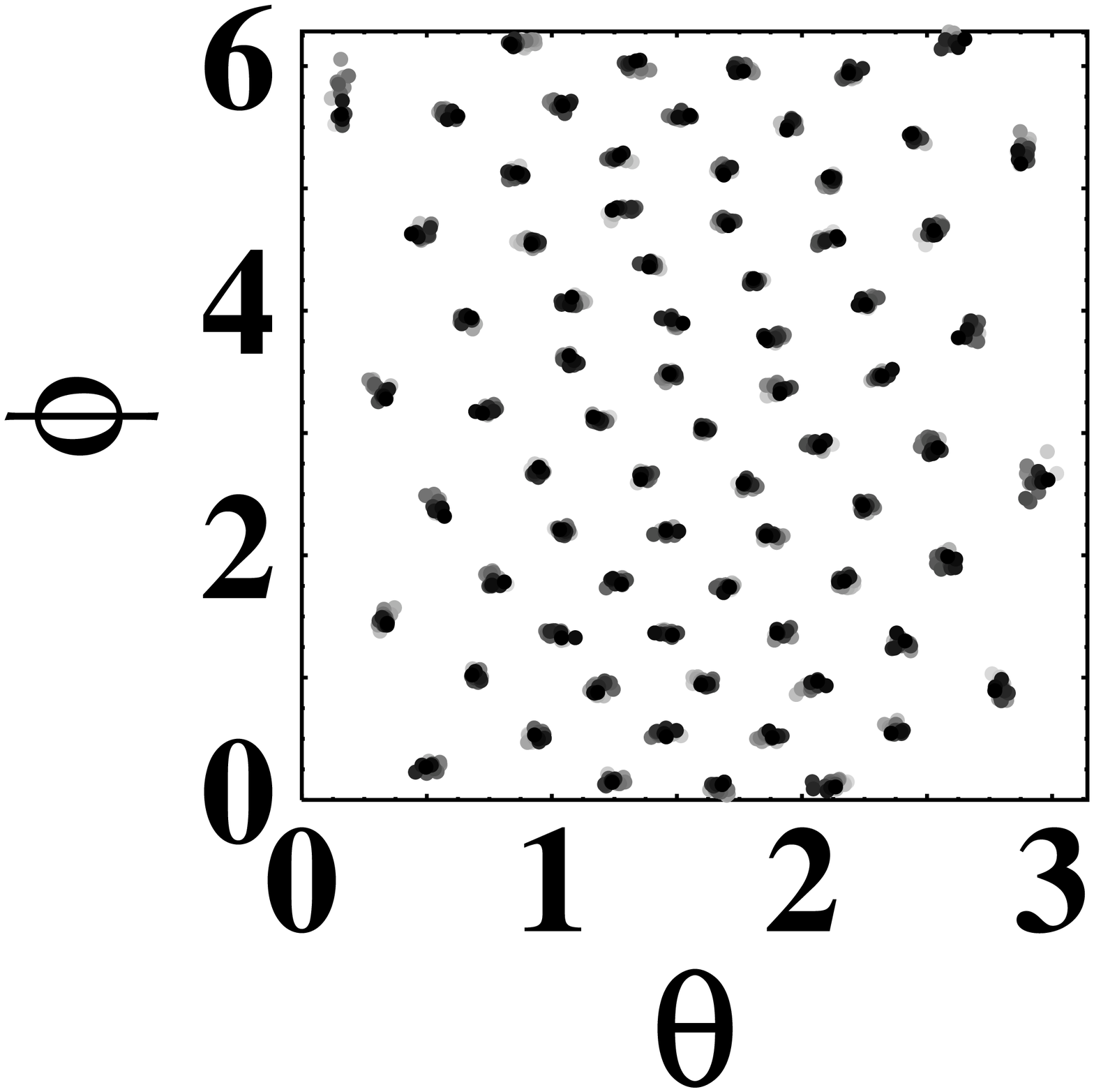}\,\,\,%
            \epsfxsize= 2.7cm\epsfbox{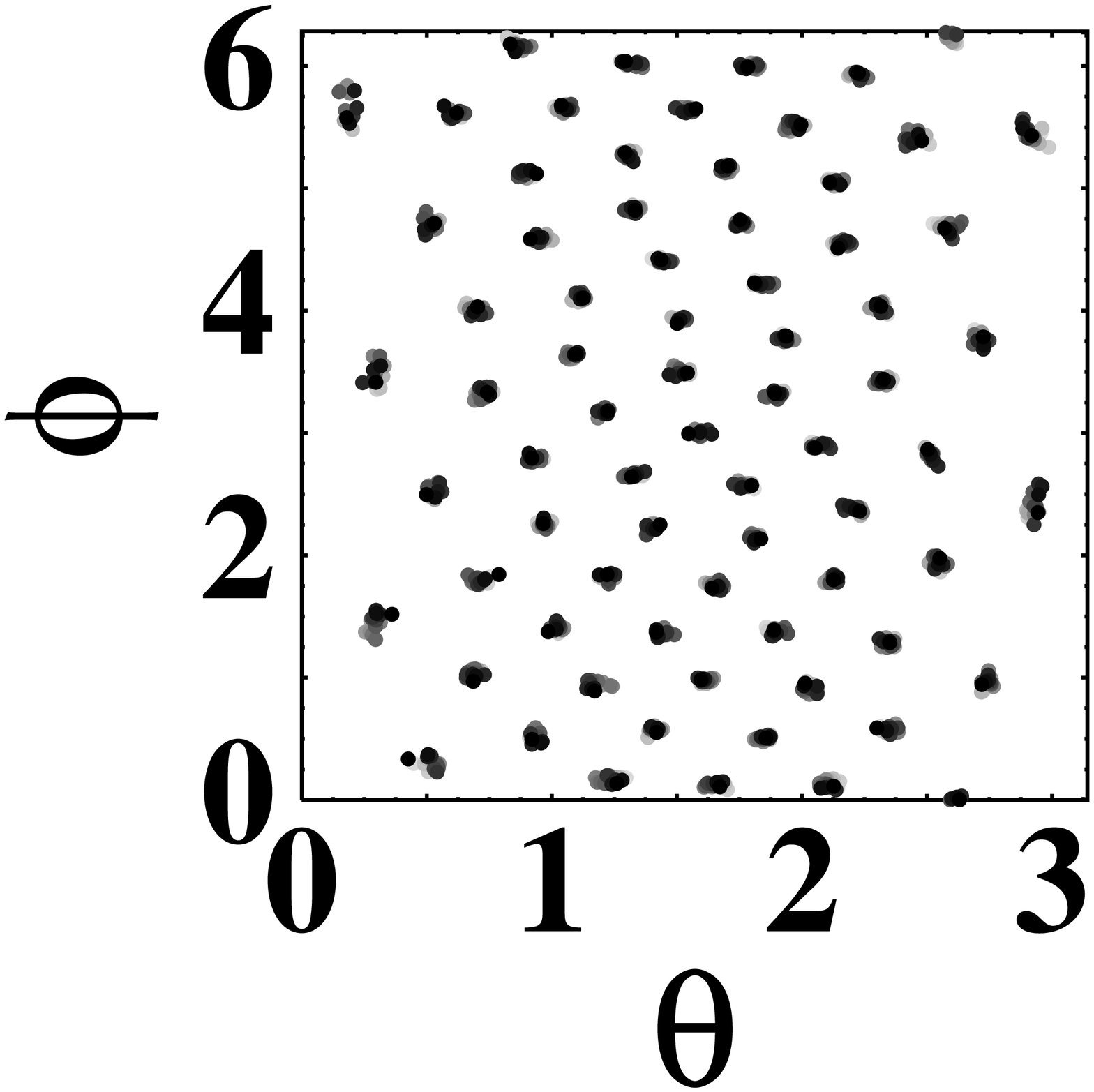}}
  \vglue 0.0em
  \centerline{\hglue 1.96cm (d) \hglue 2.35cm (e) 
              \hglue 2.36cm (f) \hglue 1.21cm}
  \vglue 0.1em
  \caption{Angular vortex-positions in all layers of the pure 
  layered-superconductor with interlayer coupling $\eta=0.14$ for 
  $(N,M)=(72,18)$ as the temperature is lowered: $\alphaT=-3,-4,-5,-6,-7,-8$. 
  [(a),(b),(c),(d),(e),(f) respectively.] Equal gray-tone represent the same 
  layer. The appearance of pancake-vortex domains signals the beginning of 
  the coupled phase [which also appears at a temperature between $\alphaT=-3$ 
  and $-4$.]}
  \label{Fig:Pancake-vortices}
\end{figure}

\begin{figure}
\centerline{\epsfxsize= 8.3cm\epsfbox{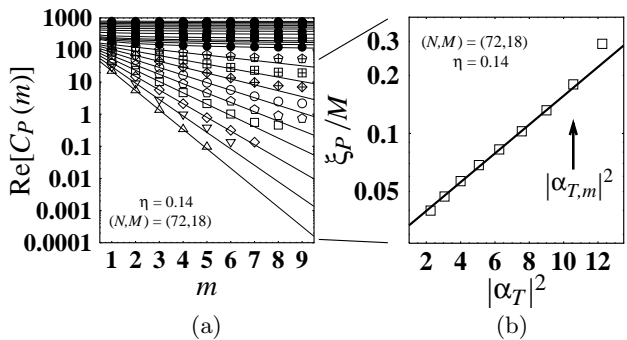}}
  \vglue 0.0em
  \centerline{\hglue 5.50cm (a) \hglue 3.53cm (b) \hglue 4.27cm}
  \vglue 0.1em
  \caption{Inter-layer phase correlations on the pure layered-HTSC as 
  temperature is lowered. (a) Log-linear plot of the inter-layer 
  phase-correlation-function, ${\rm Re}[C_P(m)]$, as a function of the 
  ``distance'' $m$ along the $c$-axis. Exponential decay is observed 
  above (open symbols) and about (crossed symbols) the transition 
  temperature, $\alphaTm$, but power-law decrease (filled symbols) below 
  the transition. In each case the mentioned curve is fitted. (b) Scaled 
  phase-correlation-length, $\xi_P/M$, vs $|\alphaT|^2$ in log-linear 
  scale for the system in (a) and also with exponential fit. [The arrow 
  marks the coupling transition temperature, which to our resolution 
  coincides with $\alphaTm$.]}
  \label{Fig:Phase-c-f}
\end{figure}

To study pancake-vortex correlations along the axis perpendicular to
the layers (the $c$-axis), we measure numerically the two-point 
correlation function
\begin{equation}
\label{eq:two-point-c-f}
C_P(m)=\frac{4\pi R^2}{Q^2}\left\langle\frac{1}{M}\sum_{n=1}^M
\overline{\psi_n^{*}({\bf r})\psi_{n+m}({\bf r})}\right\rangle.
\end{equation}
Here the overline denotes a spatial average over the surface of the
sphere of radius $R$ and the prefactor is just a ``normalization'' 
factor. We impose periodic boundary conditions on the order 
parameter: $\psi_{M+p}=\psi_p$, $p=1,2,\dots,M$. Appearance of long-range 
interlayer correlations can be observed directly from plots of the 
vortex positions in all layers as the temperature is lowered below the 
effective temperature $\alphaTm$ (see Fig. \ref{Fig:Pancake-vortices}) 
and are confirmed by the change from exponential to algebraic decay 
of the phase correlation function, ${\rm Re}[C_P(m)]$, in Fig. 
\ref{Fig:Phase-c-f}(a).
In the high-temperature regime, where the phase correlation function 
decays exponentially as ${\rm Re}[C_P(m)]\sim\exp(-m/\xi_P)$, we extract 
the phase correlation-length, $\xi_P$, and plot it (scaled by $M$) 
against $|\alphaT|^2$ (Fig. \ref{Fig:Phase-c-f}(b)). We remark that, 
analogously to $\xi_D$, Ref.~\cite{p:Moore-1997} also predicts the 
behavior $\xi_P\sim\exp(\c|\alphaT|^2)$. This form gives a consistent 
fit to the data just above $\alphaTm$ (solid line in Fig. 
\ref{Fig:Phase-c-f}(b)), but for $\alphaT\le \alphaTm$ the length scale 
$\xi_P$ grows faster than this exponential. We attribute the even 
more-rapid increase of $\xi_P$ for $\alphaT\le \alphaTm$ to an 
inter-layer coupling transition which takes place also at $\alphaTm$. 
This criterion gives us good agreement also with numerical values 
of Ref.~\cite{p:Kienappel&-1999} where the estimated coupling-transition 
temperature is about $\alphaTm=-3.5$ for $\eta=0.14$. In that investigation,
however, no freezing transition is observed!
That is the fundamental difference argued in this paper.

To summarize, we have investigated correlation functions for in-layer 
structure of the pancake-vortex system and for the coupling of the 
order parameter in different layers in the S-LD-LLL model. Results 
from MC simulations, measuring the structure factor in every layer, 
suggest that there is an effective temperature $\alphaTm$ below 
which the pancake-vortex system freezes to a crystalline phase but 
above which the system behaves as a liquid. This is signaled both by 
appearance of Bragg peaks in the vortex structure-factor $S({\bf k})$ 
and the rapid increase of the in-layer vortex-density correlation 
length, $\xi_D$. Simultaneous to this freezing transition there appears 
a coupling of the order parameter in the different layers which is 
shown directly from measurements of the vortex positions in every 
layer and is confirmed by the existence of long-ranged correlations 
in the order parameter of the distant layers. Explicitly, the 
exponential decay in correlations along the $c$-axis above the 
transition temperature $\alphaTm$ changes to algebraic decrease 
in the low temperature regime, $\alphaT < \alphaTm$. The transition 
is of first order and $\alphaTm$ depends on the interlayer 
coupling-parameter $\eta$. The LD-LLL model has shown%
~\cite{p:Hu&-1997,p:Kienappel&-1999} consistency with the experimental 
melting-curve in YBCO.

The author would like to thank Prof. Mike Moore for guidance
during his PhD and for critical reading of this manuscript. This
work was supported by CONACYT (Mexico) and partially by the
University of Manchester.

{}


\begin{thebibliography}{}
\bibitem{c:MC-steps} We employ the usual definition of one MC step as the update
                     of all $v_{n,q}$ coefficients in the system. Actually, 
                     because each $v_{n,q}$ is selected randomly to be updated; 
                     this implies that a few of them might be updated twice or 
                     so, while others might not be chosen in that MC step. 
                     This effect, however, is compensated after a large number 
                     of MC steps.
\end{thebibliography}
\end{document}